\documentclass[namedreferences]{solarphysics}
\usepackage[optionalrh,solaenum]{spr-sola-addons} 
\usepackage{graphicx}                    
\usepackage[usenames]{color}                       
\usepackage{url}                         
\usepackage{ulem} 

\newcommand{\arc}{$^{\prime\prime}$}
\newcommand{\e}{$\chi(l)$}
\newcommand{\muf}{$\langle|B_{z}|\rangle$}
\newcommand{\kl}{$\kappa_{\mathrm{L}}$}
\newcommand{\ki}{$\kappa_{\mathrm{I}}$}
\newcommand{\ks}{$\kappa_{\mathrm{S}}$}

\newcommand{\aap}{    {\it Astron. Astrophys.}}

\newcommand{\apj}{    {\it Astrophys. J.}}
\newcommand{\apjl}{    {\it Astrophys. J. Lett.}}

\newcommand{\pre}{    {\it Phys. Rev. E}}
\newcommand{\pra}{    {\it Phys. Rev. A}}
\newcommand{\solphys}{{\it Solar Phys.}}
\newcommand{\sovast}{ {\it Sov. Astron.}}

\newcommand{\memsai}{    {\it Mem. Soc. Astron. Ital.}}
\newcommand{\planss}{    {\it Planet. Space Science}}

\usepackage{solaheader}	
\newcommand{\solaaccepteddate}{15 Sep 2012}
\newcommand{\Xdoi}{10.1007/s11207-012-0140-4}	

\begin{document}

\begin{article}

\begin{opening}

\title{Instrumental and Observational Artifacts in Quiet Sun Magnetic Flux Cancellation Functions}

%
\author{A.~\surname{Pietarila}$^{1}$ \sep J.~\surname{Pietarila Graham}$^{2}$      
       }

%
\runningauthor{A. Pietarila, J. Pietarila Graham}
\runningtitle{Cancellation Functions of Quiet Sun Magnetic Flux}

%
  \institute{$^{1}$ National Solar Observatory, 950
  N. Cherry Avenue, Tucson, AZ 85719, USA
                     email: \url{apietarila@nso.edu} 
             $^{2}$Solid Mechanics and Fluid Dynamics (T-3) \& Center for Nonlinear Studies; Los Alamos National
  Laboratory MS-B258; Los Alamos NM 87545, USA
                     email: \url{jpietarilagraham@mailaps.org}
             }

\begin{abstract}
 Under the assumption that the photospheric quiet-Sun magnetic
  field is turbulent, the cancellation function has previously
  been used to estimate the true, resolution-independent mean,
  unsigned vertical flux $\langle|B_z|\rangle_{\mathrm{true}}$. We
  show that the presence of network elements, noise, and seeing
  complicate the measurement of accurate cancellation functions and
  their power-law exponents $\kappa$. Failure to exclude network
  elements previously led to too low estimates of both the
  cancellation exponent $\kappa$ and of
  $\langle|B_z|\rangle_{\mathrm{true}}$.  However, both $\kappa$ and
  $\langle|B_z|\rangle_{\mathrm{true}}$ are over-estimated  due to
  noise in magnetograms.  While no conclusive value can be derived
  with data from current instruments, our {\it Hinode}/SP results of
  $\kappa\lessapprox0.38$ and
  $\langle|B_z|\rangle_{\mathrm{true}}\lessapprox 270\,$gauss can be
  taken as upper bounds.
\end{abstract}

%

\end{opening}

%

\section{Introduction} 

 Turbulent fields possess self-similar power-laws, {\it i.e.}, they are
  fractals \cite{BrPrSe+1992,F95}.  Such power-laws ({\it i.e}, self-similarity) are
found in systems where the underlying  {physical processes} are the
same at all scales ({\it e.g.}, vortex  {or} flux tube stretching suggested by
\inlinecite{F95}).  {In these systems, the fields have the same
  degree of complexity (look the same) regardless of the observational
  resolution.  For example,} incompressible magnetohydrodynamics (MHD)
is expected \cite{I64,K65,GoSr1995}  {and found ({\it e.g.}},
  \opencite{LeBrPo+2009}; \opencite{PiGrMiPo2011}) to display
self-similar magnetic energy spectra.   {Power-law magnetic energy
  spectra are also seen in compressible, stratified MHD photospheric
  simulations \cite{Br1995} including those with realistic radiation
  and partial ionization \cite{PiGrCaSc2009,MoPiGrPr+2011}.}
 {Power-laws are found in solar observations of the line-of-sight
  magnetic energy,} kinetic energy, and even the granulation intensity
\cite{AbYu2010,AbYu2010b,GoYuCa+2010}.  {Power-laws are, of course,
  fundamental in measurements of the fractal dimension of magnetic
  structures in both observations and simulations ({\it e.g.},
  \opencite{JaVoKn2003}). Both the observations and simulations above indicate that the quiet Sun magnetic field is likely turbulent.  }

 The application of turbulent self-similarity to estimate the true mean
  unsigned vertical flux from magnetograms was introduced by
  \citeauthor{PiGrDaSc2009} (\citeyear{PiGrDaSc2009}, hereafter PGDS2009).  Define the ``net''
  flux ($\mathrm{Flux}_l$) at scale $l$ for any vertical field $f_z$
  to be the flux remaining after averaging over boxes $\mathcal A_i(l)$ of edge length
  $l$:
\begin{equation}
\mathrm{Flux}_{l} = \sum_i\bigg{|}\int_{\mathcal A_i(l)\in\mathcal
A}f_z\mathrm{d}a\bigg{|}\,,
\label{eq:netflux}
\end{equation}
where the boxes partition the total area: $\bigcup_i\mathcal
A_i(l)=\mathcal A$.  The unsigned flux of the field is given by
\begin{equation}
\mathrm{Flux}_{0} = \int_{\mathcal A}|f_z|\mathrm{d}a\,.
\label{eq:netflux}
\end{equation}

The ratio of the net flux at length scale $l$ to the true mean
unsigned flux ({\it i.e.}, flux {\it cancellation}) is called the
partition function
\begin{equation}
\chi(l) = \frac{\mathrm{Flux}_{l}}{\mathrm{Flux}_{0}}\,,
\label{eq:partition}
\end{equation}
which follows a power-law for self-similar fields,
\begin{equation}
\chi(l)\propto l^{-\kappa}\,,
\label{eq:kappa}
\end{equation}
where $\kappa$ is the cancellation exponent \cite{ODS+92}.  For a
completely self-similar field, knowledge of net flux at any scale and
of the power-law scaling exponent implies knowledge of net flux at all
scales and, hence, the unsigned flux.  The two extreme examples are a
unipolar field, for which the net flux always equals the unsigned flux
($\kappa=0$), and a random field.  For a random field, the observed
net flux equals the unsigned flux times the ratio of the noise
correlation length to the diameter of the resolution element
($\kappa=1$).  See the Appendix for a mathematical derivation of the
random noise case and of the general formula, Equation
(\ref{eq:extrapolate}), to determine the unsigned flux for a
completely self-similar field given $\kappa$ and the correlation
length below which the field becomes smooth.

Under the assumption that the quiet-Sun magnetic field is
turbulent, the theoretical framework in the Appendix applies. This
might suggest that the unsigned quiet-Sun flux could be deduced from
Zeeman-based instruments.  Such studies have previously been
made (PGDS2009; \opencite{St2011}). PGDS2009 observed a power-law in $\chi(l)$ (which
they dubbed the ``cancellation function'' when applied to a
magnetogram) down to $\approx200\,$km with $\kappa=0.26$ in {\it
Hinode} Spectro-Polarimeter (SP; \opencite{HinodeSOT}) data.  Note
that power-laws for $\chi(l)$ have previously been seen in
simulations of the electric currents \cite{SVCN+02,PGMP05} and in solar
observations of current-helicity \cite{SoVaAbCa+2003b,Ab2003,SVCV+04,AbYu2010}.
PGDS2009 used Equation (\ref{eq:extrapolate}) to estimate the net
flux at a scale of $800\,$m to be $\approx 50\,$G.  Turbulent
power-laws ({\it e.g.}, energy spectra and $\chi(l)$) extend only down to the
dissipative range, but the magnetic diffusion scale is expected to
be significantly smaller than $800\,$m.  Thus, the extrapolation was
taken as a lower bound, $\mathrm{Flux}_0>50\,$G (gauss).  This result was
found to be in agreement with a separate extrapolation from
radiative MHD simulations.

Other estimates of $\kappa$ can be made.  \inlinecite{LiKuSoNa+2008}
found a 30\% increase in flux changing from a resolution of 1\arc\ to
0.33\arc\ which indicates $\kappa=0.24$.  \inlinecite{SoBaDa+2010}
found a factor of 3.7 more flux on doubling of resolution, indicating
$\kappa=1.9$.  Using many instruments with different spatial
resolutions, \inlinecite{SaAlMaGo2011} fit $\mathrm{Flux}_l \propto
l^{-1}$, {\it i.e.}, $\kappa$=1. The data set used by PGDS2009 was recalibrated by \inlinecite{St2011}
leading to more pronounced high field strength tails in the flux
probability distribution. The power law from the recalibrated
magnetogram yielded a value of $\kappa$ that is half of the value
derived by PGDS2009. In addition to the SP magnetogram \inlinecite{St2011}
computed cancellation functions from a set of {\it Hinode} Narrow-band Filter Imager (NFI) Na {\sc i}
$D_{1}$ magnetograms. The resulting $\kappa$ was found to be 0.127,
{\it i.e.}, very similar to the value for the recalibrated SP magnetogram. In order to agree with Hanle-based magnetic field measurements, this cancellation function needs to be extrapolated down to a spatial scale of 10\,--\,100 m.

Another application for using the fractal nature of magnetic fields is
in flare predictions. Previous works (\opencite{SoVaAbCa+2003b}, \citeyear{SVCV+04}; \opencite{AbYu2010}) have shown it to be
promising. However, as in the case of cancellation functions, the data-sample sizes used to study the suitability of the method have been
limited. Recently \inlinecite{Georgoulis2011} has shown how previous
analyses may have been misleading: A statistical comparison of flaring and
non-flaring active regions found that the fractal properties of the line-of-sight magnetic field cannot
distinguish flaring and non-flaring active regions. Many of the previous works, however, used current helicity, not magnetic flux, for the analysis and therefore the conclusions of \inlinecite{Georgoulis2011} may not apply to them.  

Prompted by the results of PGDS2009, \inlinecite{St2011}, and
\inlinecite{Georgoulis2011}, we suggest that measurements of the
cancellation function do not confront the reality of the
observational data.  Such confrontation is the aim of this paper.
We analyze the statistics of quiet Sun cancellation functions using
magnetograms from three different instruments.  We test if the
cancellation exponent $\kappa$ is robust within and among the
instruments to point out observational artifacts in the measurement of
$\kappa$ and to determine if any conclusive values or bounds can be
made with existing data.

\section{Data and Methods}

We use altogether $\approx$800 magnetograms from the Helioseismic
and Magnetic Imager (HMI; \opencite{HMI}) on the {\it Solar Dynamics Observatory} (SDO) satellite, Michelson Doppler
Imager (MDI; \opencite{MDI}) on the {\sl Solar and Heliospheric Observatory} (SOHO) sarellite, and {\it Hinode} SP. The largest data
set, 700 HMI magnetograms, allows us to characterize the variation
within a single instrument. Data from MDI and SP allow for
inter-instrument comparisons and  {for} identification of possible
instrumental artifacts as well as  {for determination of the
effects} of spatial resolution and magnetogram noise.

The HMI data consist of daily (from end of March 2010
to mid-June 2011) 4 s integration magnetograms taken around 12:00 UT. We analyze a $501\times 501$ pixel ($\approx 253$ \arc $\times 253$ \arc) area around
the disk center. For MDI, we use $301 \times 301$ pixel ($\approx 182$ \arc $\times 182$ \arc)
sub-regions, excluding active regions, in high resolution magnetograms
(level 1.8) near the solar disk center. Note that a calibration
coefficient ({\sf BSCALE}=2.81) is applied to the MDI magnetograms to convert
from counts to gauss leading to the magnetograms being quantized in units
of {\sf BSCALE}.

The SP magnetograms are the longitudinal magnetic field measurements
in Level 1D data with exposure times varying between 1.6 and 12
s. Both $\approx$0.16\arc\ and $\approx$0.32\arc\ pixel size magnetograms are
included in the data set. We choose observations of quiet Sun taken
near the disk center. The size of SP magnetograms ranges from $\approx$
$400 \times 400$ to $1000 \times 1000$ pixels ($\approx 64$ \arc $\times 64$ \arc--160 \arc $\times 160$ \arc).

Cancellation functions [\e] are computed for each
magnetogram using the Monte Carlo technique
of \inlinecite{CaLaRu+1994}.  Linear fits  {to $\chi(l)$ in log-log
space are made} for three different spatial scales: small scale $<$2\arc\  {(exponent} \ks),
intermediate scale 2\,--\,5\arc\ (\ki) and large scale 2\,--\,9 \arc\
(\kl). While the physical scales are the same for all instruments, the
number of pixels for fitting the exponent varies due to the
instruments' different pixel sizes (HMI $\approx$0.5\arc, MDI
$\approx$0.6\arc, SP $\approx$0.16\arc or $\approx$0.32\arc). {The spatial scales were chosen to be sensitive to the scales most affected by noise (\ks) and scales not dominated by noise, but still below the spatial scale of the network (\ki\ and \kl). The overlap of \ki\ and \kl\ ensures that enough data points (14 for HMI, 11 for MDI and 24 or 47, depending on pixel size, for SP) are included in fitting \kl\ and that the results are not strongly sensitive to the upper cut-off (5\arc\ or 9\arc) of the fitted spatial scales.} 

Additionally, the
mean unsigned flux {\muf} and flux imbalance {$\langle B_{z}\rangle$/\muf} are
computed for each magnetogram.

\section{Results}

\subsection{Exponents and Effects of Network}

\begin{figure*}
\includegraphics[width=12.cm]{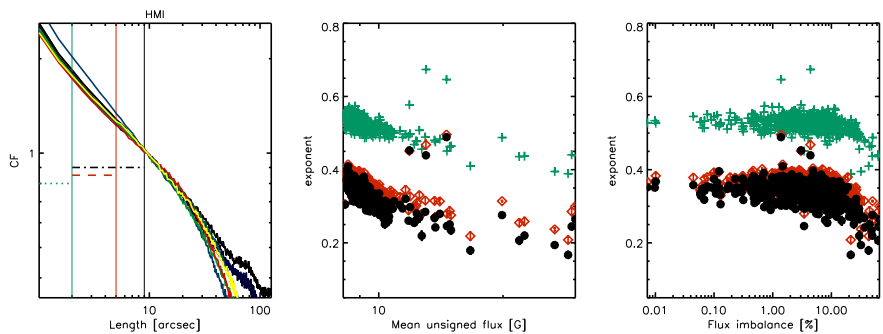}
\caption{HMI cancellation functions. Left: Cancellation functions for
selected magnetograms (in various colors) normalized to unity at
9\arc. Vertical and horizontal lines mark ranges used for fitting
the exponents (\ks\ in green dotted line and crosses, \ki\ in red dashed line and diamonds, and \kl\ in black dash-dotted line and circles). The
same colors and symbols are used in all plots to differentiate between the
spatial scales. Middle: Cancellation exponents as a function of \muf. Error
bars show 1$\sigma_{\mathrm {fit}}$. Right:
Exponents as a function of flux imbalance. }
\label{fig:hmi-all}
\end{figure*}

The results of the analysis on the HMI cancellation function are
summarized in Figure \ref{fig:hmi-all} and Table
\ref{table:hmi-mdi}. Compared to large spatial scales ($>$10\arc, for
which we do not  {fit a power-law), the exponents} at the smaller
scales are fairly similar: \ki\ and \kl are nearly identical, which
indicates that we are fitting a genuine power-law at these scales (
2\arc to 9\arc). \e\ turns up at the small scales as is demonstrated
by \ks\ having $\approx$30\% higher values than \ki\ and \kl. The
increase is partially due to contributions from noise becoming more
dominant (see Section \ref{sec:noise} and note that
$\kappa_{\mathrm{noise}}=1$; \inlinecite{SVCN+02}  {and the Appendix). Note that there
is both scatter (standard deviation up to 6\% of the mean) and
systematic differences between the exponents of power-law fit to different
magnetograms.}

{All exponents decrease} as a function of \muf,  {\muf$\equiv \mathrm{Flux}_l$ for $l=1\,$pixel.} The more
flux there is in the magnetogram, the smaller  {each} $\kappa$ is. A linear
fit of $\kappa$ as a function of $\log$\muf\ shows that the decrease
is strongest for \kl\ and weakest for \ks. No correlation is found for
$\kappa$ and the global (full-disk) unsigned flux, which reflects the overall
activity present on the solar disk at a given time: The
cancellation functions reflect the local magnetic environment, whose
fractal properties do not appear to be influenced by the global
field.
$\kappa$ does not depend on the flux imbalance at small imbalances
(below $\approx$10\%). For imbalances greater than $\approx$10\%\ $\kappa$
decreases with increasing imbalance on all spatial scales.

\begin{figure*}
\includegraphics[width=10cm]{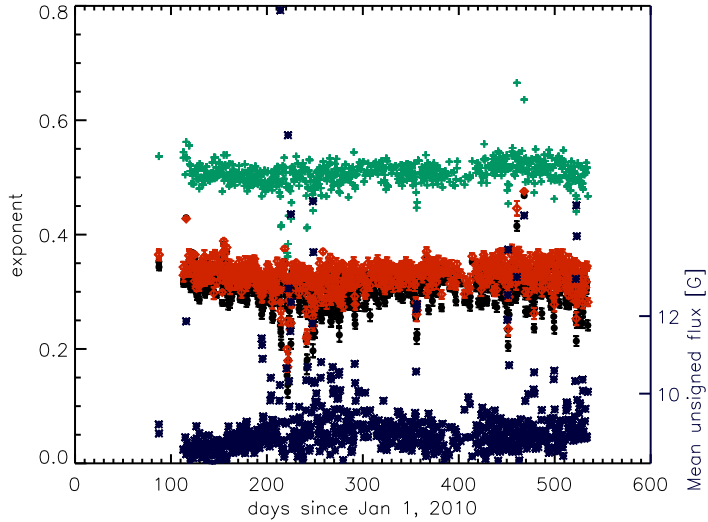}
\caption{HMI cancellation exponents as a
function of time. \ks\ is shown in green crosses, \ki\ in red diamonds, and \kl\ in black circles. \muf is shown in dark blue $\ast$ (y-axis on right). }
\label{fig:hmi-ts}
\end{figure*}

The exponents show no strong temporal {evolution} or
indications of the exponent changing as solar activity increases
(Figure \ref{fig:hmi-ts}). Since the activity in the rising phase of
the solar cycle emerges at high latitudes, it is not surprising that no significant change in the exponent is seen near disk center. 

\begin{figure*}
\includegraphics[width=12.cm]{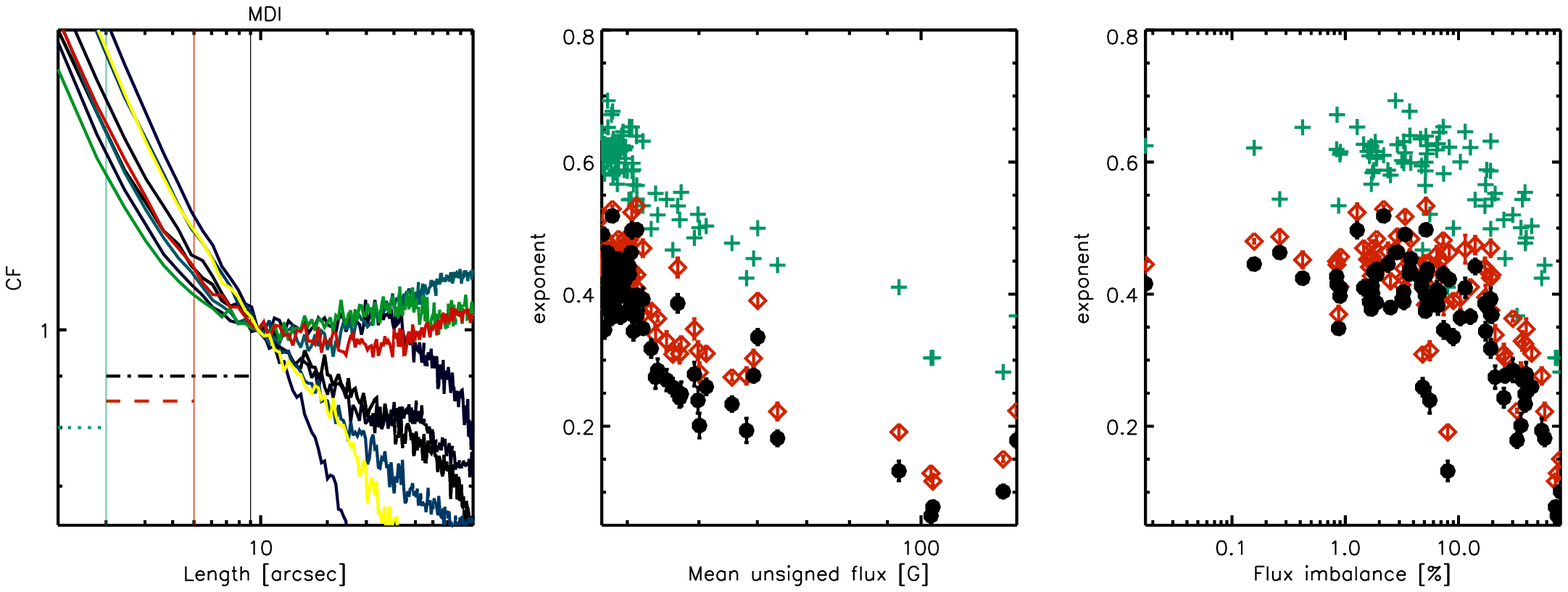}
\caption{MDI cancellation functions. Left: Cancellation functions for
selected magnetograms (in various colors) normalized to unity at
9\arc. Vertical and horizontal lines mark ranges used for fitting
the exponents (\ks\ in green dotted line and crosses, \ki\ in red dashed line and diamonds, and \kl\ in black dash-dotted line and circles). The
same colors and symbols are used in all plots to differentiate between the
scales. Middle: Cancellation exponents as a function of \muf. Error
bars show 1$\sigma_{\mathrm {fit}}$. Right:
Exponents as a function of flux imbalance.}
\label{fig:mdi-all}
\end{figure*}

The results for MDI are shown in Figure \ref{fig:mdi-all} and Table
\ref{table:hmi-mdi}. They are qualitatively similar to HMI, confirming
that the trends (\muf, flux imbalance) are of solar, not instrumental,
origin. The exponents are larger and have more scatter  {(standard
deviation is 9\% of the mean).} Also the dependence of $\kappa$ on
\muf\ is stronger.

\begin{table}
\caption{ {Power-law fits to $\chi(l)$ for HMI, MDI, and SP magnetograms.}}
\label{table:hmi-mdi}
\label{table:sp}
\begin{tabular}{lllll}
\hline
& Mean & Standard deviation & c\footnotemark[1] & f$^{\mathrm{1}}$ \\
\hline%
HMI\footnotemark[2]   &       &        &             &              \\
\ks    & 0.53  & 0.017 & 0.74 (0.01) & -0.22 (0.01) \\
\ki    & 0.36  & 0.020 & 0.61 (0.01) & -0.26 (0.01) \\
\kl    & 0.35  & 0.022 & 0.66 (0.01) & -0.34 (0.01) \\
MDI$^\mathrm{2}$    &       &        &             &              \\
\ks    & 0.61  & 0.053 & 1.62 (0.06) & -0.64 (0.03) \\
\ki    & 0.45 &  0.033 & 1.47 (0.07) & -0.65 (0.04) \\
\kl    & 0.42 &  0.036 & 1.51 (0.08) & -0.70 (0.05) \\
SP    &       &        &             &              \\
$t_{\mathrm{exp}} <$2 s & & & &  \\
\ks & 0.30 &0.020 &-- & -- \\
\ki & 0.28 &0.020 &-- & -- \\
\kl & 0.29 &0.014 &-- & -- \\
$t_{\mathrm{exp}} >$2 s & &  & &\\
\ks & 0.24 &0.0063 &-- & -- \\
\ki & 0.26 &0.014 &-- & -- \\
\kl & 0.28 &0.032 &-- & -- \\
\hline
\end{tabular}
\newline
\footnotetext{1}{$^{\mathrm{1}}\kappa=c$+$f$*$\log$(\muf) $\sigma_{\mathrm {fit}}$ is given in parentheses.}\newline
\footnotetext{2}{$^{\mathrm{2}}$Only magnetograms with flux imbalances below 5\% are
included.}  
\end{table}

\begin{figure*}
\includegraphics[width=12.cm]{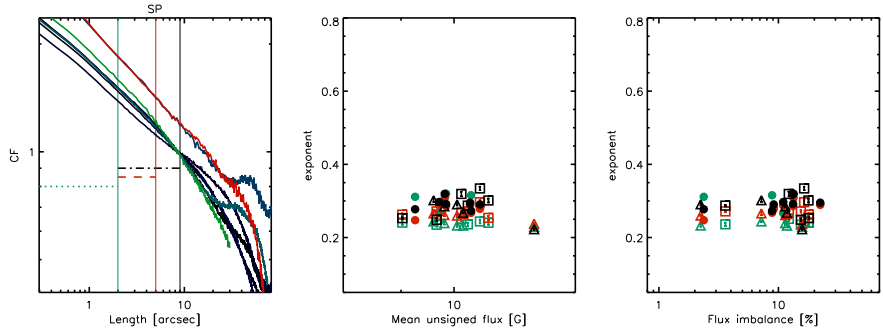}
\caption{SP cancellation functions. Left: Cancellation functions for
selected magnetograms (in various colors). The cancellation functions with pixel size 0.32\arc are normalized to unity at 18\arc and pixel size 0.16\arc at 9\arc. Vertical and horizontal lines mark ranges used for fitting
the exponents (\ks\ in green dotted line, \ki\ in red dashed line, and \kl\ in black dash-dotted line). The
same colors are used in all plots to differentiate between the
spatial scales. Middle: Cancellation exponents as a function of \muf. Error
bars show 1$\sigma_{\mathrm {fit}}$. Circle symbols are rasters
with exposure times less than 2 s, diamonds are exposure times between 2 and 5 s, and squares are exposure times longer than 5 s. Right:
Exponents as a function of flux imbalance.   }
\label{fig:sp-all}
\end{figure*}

The SP magnetograms (Figure \ref{fig:sp-all} and Table \ref{table:sp})
are a mixture of data with different pixel sizes and exposure times. A
comparison of \ks\ for magnetograms with exposure times less than two
seconds and ones with above, shows that \ks\ is larger for shorter
exposure (and thus noisier) magnetograms. The difference is smaller
for \ki\ and \kl. This is to be expected, if the difference is due to
noise, which is more dominant at small scales (see Section
\ref{sec:noise}). The measured exponents are similar to $\kappa$ found
by PGDS2009 who measured $\kappa$=0.26  {from a single
magnetogram}. Regardless of exposure time, all the SP exponents are
smaller than MDI and HMI exponents. No dependence of $\kappa$ on
\muf\ or flux imbalance is seen. The sample, however, is too small to
establish this. Note that for the magnetograms with the longest
exposure times a mixture of spatial and temporal cancellation may
occur. The time to raster 2\arc\ varies strongly depending on the
exposure time and step size: For exposure times of 1.6 and 4.8 s
the rastering times are well below the granulation turnover time
scale. For the exposure time of 8 s and above the rastering time is
over 100 s.

The dependence of MDI and HMI cancellation exponents on \muf\ and flux
imbalance suggest that the presence of magnetic network elements in
the magnetograms influences the cancellation exponents. (Magnetic
network flux is stronger than the intra-network, the network is mostly unipolar, and
occurs at scales of a few tens of arcseconds.) To  {model} the
effect on $\chi(l)$ of a large (active-region-remnant or network-like)
unipolar flux region, our observation region $\mathcal{A}$ is
decomposed into two sub-regions: $\mathcal{B}$ (containing
intra-network quiet-Sun in which a scaling $\chi(l)$ would be found)
and the unipolar flux concentration $\mathcal{C}$ (over which
$\int_{\mathcal C}|B_z|\mathrm{d}a = \big{|}\int_{\mathcal C}B_z\mathrm{d}a\big{|}$,
{\it i.e.}, there is no cancellation). Our goal is to find the cancellation
function [$\chi'(l)$] over the entire region $\mathcal{A} =
\mathcal{B}\cup\mathcal{C}$. We consider only scales $l\ll$ the
size of the unipolar flux concentration. So, a given square $\mathcal
A_i(l)$ is either in $\mathcal{B}$ or $\mathcal{C}$,
\begin{eqnarray}
\chi'(l) = \frac{\sum_{\mathcal A_i(l)\in\mathcal
B}\bigg{|}\int_{\mathcal A_i(l)}B_z\mathrm{d}a\bigg{|}+\sum_{\mathcal
A_i(l)\in\mathcal C}\bigg{|}\int_{\mathcal
A_i(l)}B_z\mathrm{d}a\bigg{|}}{\int_{\mathcal B}|B_z|\mathrm{d}a + \int_{\mathcal
C}|B_z|\mathrm{d}a}\,.
\end{eqnarray}
Using $\sum_{\mathcal A_i(l)\in\mathcal C}\bigg{|}\int_{\mathcal
 A_i(l)}B_z\mathrm{d}a\bigg{|}=\int_{\mathcal C}|B_z|\mathrm{d}a$ and defining
 \newline$P=\int_{\mathcal C}|B_z|\mathrm{d}a/\int_{\mathcal B}|B_z|\mathrm{d}a$, the ratio of
 mean unsigned flux in the unipolar flux concentration compared to
 the rest of the region, we find
\begin{eqnarray}
\chi'(l) = \frac{\chi(l)+P}{1+P}\,.
\label{eq:model}
\end{eqnarray}
Assuming that the intra-network quiet-Sun flux is completely balanced,
flux imbalance (FI) is given by
\begin{equation}
\mathrm{FI} = \frac{\bigg{|}\int_{\mathcal A}B_z\mathrm{d}a\bigg{|}}{\int_{\mathcal
A}\big{|}B_z\big{|}\mathrm{d}a} = \frac{P}{1+P}\,.
\label{eq:fip}
\end{equation}

\begin{figure*}
\includegraphics[width=10cm]{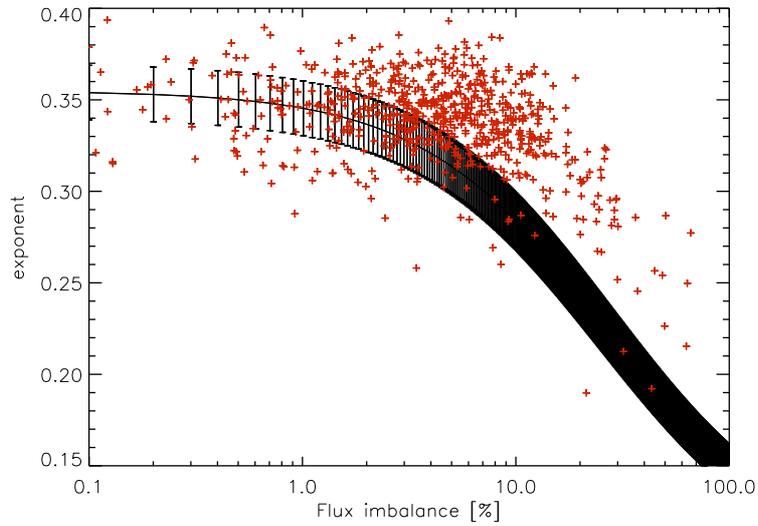}
\caption{Effect of flux imbalance. Synthetic cancellation
exponents for intermediate scale {\ki} as a function of flux
imbalance. Error bars are 1$\sigma_{\mathrm{\mathrm {fit}}}$.
Measured HMI \ki\ are shown as red crosses.}
\label{fig:unipolar}
\end{figure*}

{We take $\chi(l)$ from}
the HMI magnetogram with the
smallest flux imbalance
{and employ Equations (\ref{eq:model}) and (\ref{eq:fip})
to plot synthetic $\chi'(l)$ versus flux
imbalance in Figure \ref{fig:unipolar}.} The synthetic exponents as a
function of flux imbalance mimic the MDI and HMI observations: Small
increases in the imbalance do not alter the exponent, while imbalances
above $\approx$10\%\ lead to a decreasing $\kappa$. The model explains
the observed dependence of $\kappa$ on flux imbalance as a consequence
of unipolar network magnetic fields altering the cancellation
function. Note that the dependence of the exponents on \muf\ applies
also to magnetograms with very small imbalances. Small imbalances
imply that the positive and negative network concentrations balance
out each other, not that there is only a small amount of network flux
present.

\subsection{Intra-Network Exponents}

\begin{figure*}
\includegraphics[width=12.cm]{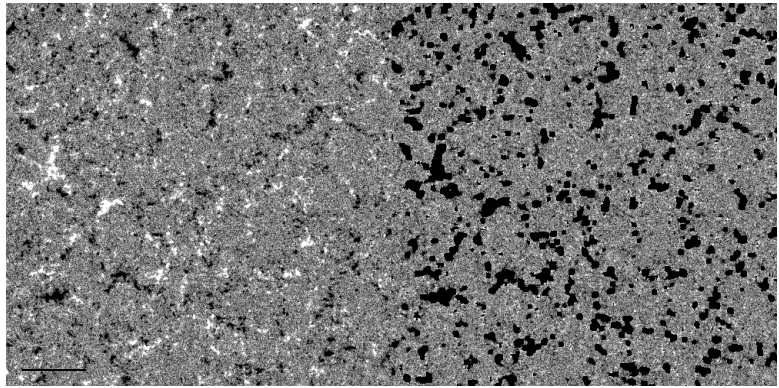}
\caption{Masking of network. Left: Original HMI magnetogram with
network elements. Right: Same magnetogram with network elements masked
out. The color scales saturate at $\pm$25 G. The horizontal line in
the bottom left corner shows the spatial scale of 50\arc. }
\label{fig:mask}
\end{figure*}

Since the cancellation exponents are affected by the presence of
network elements, a combination of smoothing and thresholding  {can be}
applied to mask out all the network pixels in the magnetograms prior
to computing the cancellation functions. An example of a mask applied
to an HMI magnetogram is shown in Figure \ref{fig:mask}. To test how
the masking affects the measured exponents, we apply a network mask
to a synthetic magnetogram consisting of pure noise ($\kappa_{\mathrm{noise}}=1$) and compare the
exponents from the masked and unmasked magnetograms: The {effect on} the exponent is smaller than the error in the fit.

\begin{figure*}
\includegraphics[width=12.cm]{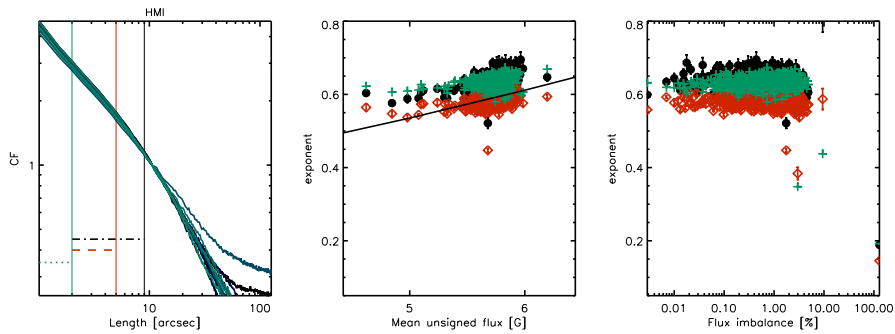}
\caption{ HMI cancellation functions for intra-network (network masked out) magnetograms. Left: Cancellation functions for
 selected magnetograms (in various colors) normalized to unity at
 9\arc. Vertical and horizontal lines mark ranges used for fitting
 the exponents (\ks\ in green dotted line and crosses, \ki\ in red dashed line and diamonds, and \kl\ in black dash-dotted line and circles). The
 same colors and symbols are used in all plots to differentiate between the
 spatial scales. Middle: Cancellation exponents as a function of \muf. Solid black line shows linear fit, slope 0.3,  of $\kappa$ vs. $\log$\muf. Error
 bars show 1$\sigma_{\mathrm {fit}}$. Right:
 Exponents as a function of flux imbalance.  {The systematic dependence is
seen to be effectively removed.}}
\label{fig:hmi-qs}
\end{figure*}

\begin{figure*}
\includegraphics[width=12.cm]{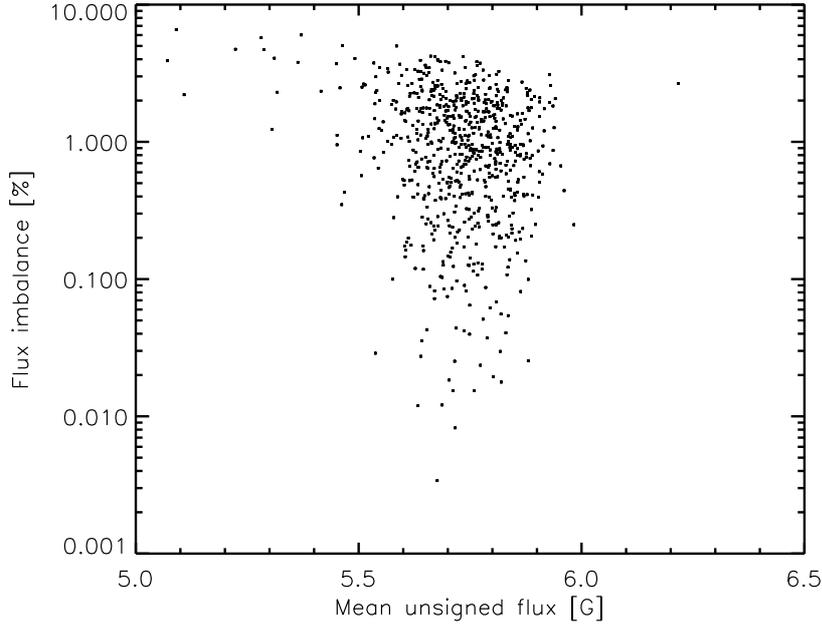}
\caption{Flux imbalance in masked HMI magnetograms vs. \muf. }
\label{fig:for-j}
\end{figure*}

Masking out the network pixels in the HMI magnetograms to measure the
``true'' intra-network cancellation exponents (Figure \ref{fig:hmi-qs}
and Table \ref{table:inw}) leads to significantly  higher values for
\kl\ and \ki. $\kappa$ still depends on \muf, but in the opposite sense  {than}
in the unmasked magnetograms: In the intra-network magnetograms
$\kappa$ increases with increasing \muf. This demonstrates that the
decrease of $\kappa$ with increasing \muf\ in the unmasked
magnetograms was due to the presence of the network elements. A linear
fit of intra-network magnetogram $\kappa$ vs.\, $\log$(\muf) gives a
slope of $\approx$0.3. This increase of $\kappa$ with \muf\ can be
 understood from Equation (\ref{eq:model}).  An increase in flux imbalance
 means more pixels that are never canceled out at any scale, a
 flattening of the cancellation function, and a decrease in $\kappa$.
 A magnetogram with very weak signal can be out of balance from a
 small number of moderately strong pixels of the same sign.  In
 fact, we see in Figure \ref{fig:for-j} that nearly all magnetograms
 with over 1\% flux imbalance, have \muf$<\,5.5\,$G.
 The weakest magnetograms have greater flux imbalances and,
 consequently, smaller $\kappa$.

\begin{figure*}
\includegraphics[width=12.cm]{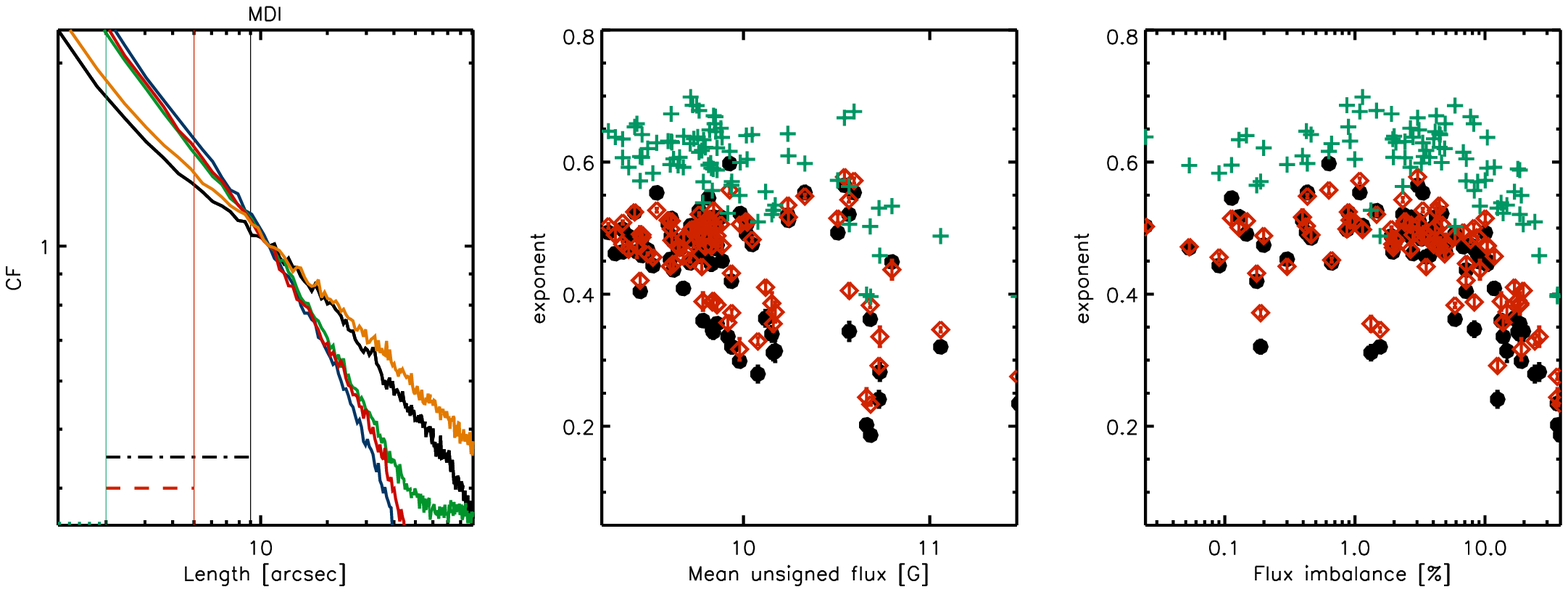}
\caption{MDI intra-network (network masked out) cancellation functions. Left: Cancellation functions for
 selected magnetograms (in various colors) normalized to unity at
 9\arc. Vertical and horizontal lines mark ranges used for fitting
 the exponents (\ks\ in green dotted line and crosses, \ki\ in red dashed line and diamonds, and \kl\ in black dash-dotted line and circles). The
 same colors and symbols are used in all plots to differentiate between the
 scales. Middle: Cancellation exponents as a function of \muf. Error
 bars show 1$\sigma_{\mathrm {fit}}$. Right:
 Exponents as a function of flux imbalance. }
\label{fig:mdi-inw}
\end{figure*}

\begin{figure*}
\includegraphics[width=12.cm]{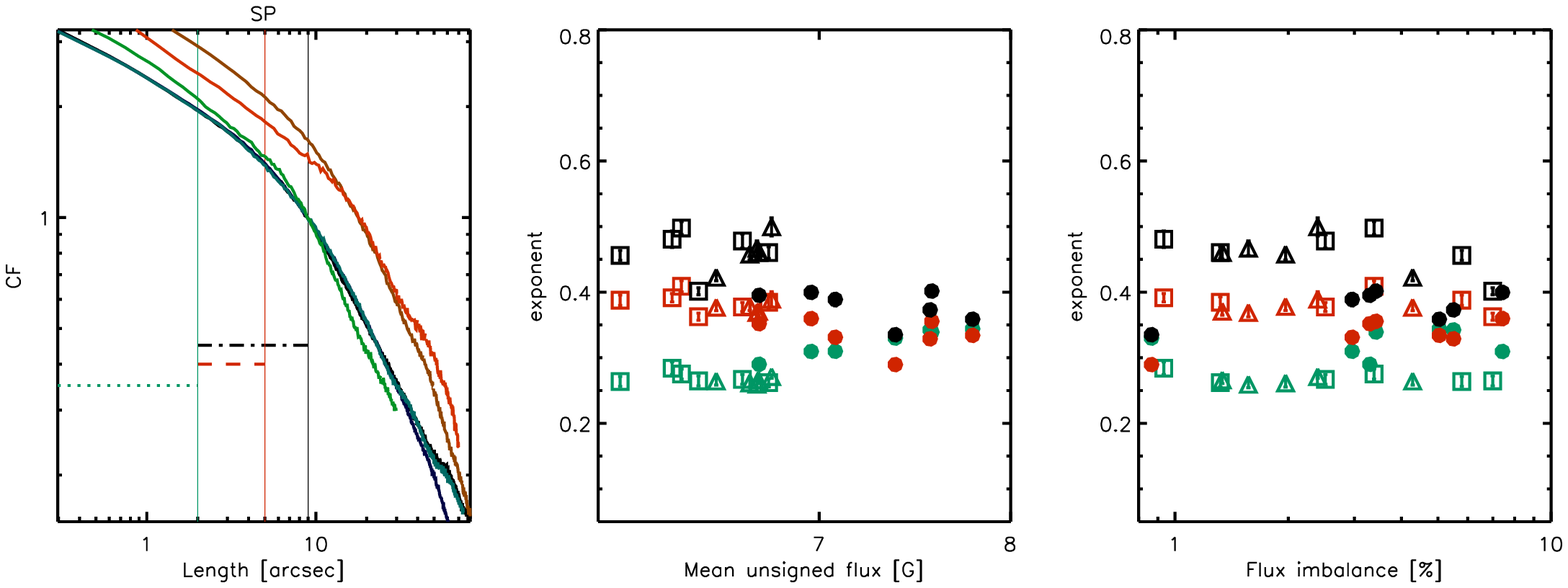}
\caption{SP intra-network
 (network masked out) cancellation functions. Left: Cancellation
 functions for selected magnetograms (in various colors). The
 cancellation functions with pixel size 0.32\arc are normalized to
 unity at 18\arc and pixel size 0.16\arc at 9\arc. Vertical and
 horizontal lines mark ranges used for fitting the exponents (\ks\ in
 green dotted line, \ki\ in red dashed line, and \kl\ in black
 dash-dotted line). The same colors are used in all plots to
 differentiate between the spatial scales. Middle: Cancellation
 exponents as a function of \muf. Error bars show 1$\sigma_{\mathrm
   {fit}}$. Circle symbols are rasters with exposure times less than
 2 s, diamonds are exposure times between 2 and 5 s, and
 squares are exposure times longer than 5 s.  {Most of the
   scatter is due to differing exposure times.} Right: Exponents as a
 function of flux imbalance. }
\label{fig:sp-qs}
\end{figure*}

The network
masking of MDI magnetograms was not entirely successful and effects of
network are still visible in the \muf\ and flux imbalance plots. Compared to HMI, the effect of masking the magnetograms is small in
MDI (Figure \ref{fig:mdi-inw} and Table \ref{table:inw}). The
modest change may also be partially due to the network
vs. intra-network contrast being smaller in MDI than HMI. The
intra-network SP exponents (Figure \ref{fig:sp-qs} and Table
\ref{table:inw2}) are increased on all scales. The change is largest
for longer exposures and large spatial scales. Since the longer
 exposure times have less noise and their $\kappa$ values increase the most after removal of the network, the increase in
 $\kappa$ cannot be due to noise alone.

\begin{table}
\caption{ {Power-law fits} for intra-network  {(network masked
   out) cancellation functions for HMI, MDI, and SP magnetograms.}}
\label{table:inw}
\label{table:inw2}
\begin{tabular}{lll}
\hline
 & Mean & Standard deviation \\
\hline
HMI &      &        \\
\ks & 0.63 & 0.024\\
\ki & 0.58 & 0.024 \\
\kl & 0.64 & 0.084\\
MDI\footnotemark[1] &      &        \\
\ks & 0.59 &  0.055\\
\ki & 0.46 &  0.077\\
\kl & 0.46 &  0.089\\
SP  &      &        \\
$t_{\mathrm{exp}} <$2s & & \\
\ks & 0.32 & 0.020\\
\ki & 0.32 & 0.017\\
\kl & 0.38 & 0.017\\
$t_{\mathrm{exp}} >$2s & & \\
\ks & 0.28 & 0.0063\\
\ki & 0.38 & 0.014\\
\kl & 0.48 & 0.036\\
\hline
\end{tabular}
\newline
\footnotetext{1}{$^1$Only magnetograms with no network remnants are included.}
\end{table}

\subsection{Effect of Noise and Seeing}
\label{sec:noise}

\begin{figure*}
\includegraphics[width=12.cm]{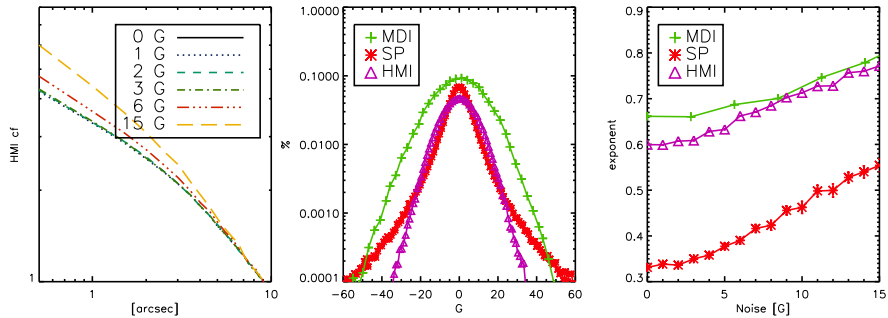}
\caption{Noise and \e. Left: HMI intra-network cancellation functions
with increasing artificial noise (see figure legend). Middle:
Histograms of intra-network  {magnetic flux.}  Right: Small-scale exponents as a function of
added noise. For the SP magnetogram we use a raster taken with a 1.6
s exposure time.}
\label{fig:noise}
\end{figure*}

An examination of Tables \ref{table:sp} and \ref{table:inw} suggests that the cancellation
exponent is larger for noisier instruments and shorter exposures, {\it
i.e.}, for noisier magnetograms.  We now demonstrate that this is
the case by artificially adding noise to our magnetograms.   {The}
first panel in Figure \ref{fig:noise} shows how an HMI cancellation
function changes as random-distributed noise is incrementally added to
the magnetogram prior to computing \e. Adding noise changes first the
smallest scales. As the amount of noise increases the scales affected
by noise  {also} increase.   {For} HMI adding 1 G (random
numbers with a mean of zero and a standard deviation of one) of noise
does not change the cancellation function noticeably and only a small
difference is seen for 3 G. In contrast,  {adding noise at the 6
and 15 G levels, respectively, significantly alters $\chi(l)$ on
increasing spatial scales.}  That no change is visible for $3\,$G, gives us
a measure of noise in the HMI magnetogram: between 3 and $6\,$G.

The remaining panels in Figure \ref{fig:noise} show how the small
scale cancellation exponents for the different instruments react
to increased noise. (Noise was added in units of counts (1
count\,=\,2.81 G) to the MDI magnetograms.) All of the \ks, as a function of noise curves, have the same general
shape (rightmost panel in Figure \ref{fig:noise}): They start with a
plateau and after an instrument-specific threshold (magnetogram noise
level) is reached, \ks\ begins to increase linearly as a function of
added noise. The nominal noise values (upper
limits) of the different instruments can be defined as the full width at half maximum of Gaussian fits to the  {magnetic flux} histograms. The widths are
31.6 G for MDI, 21.1 G for HMI, and 13.5 G for SP.
In the linear increase regime the increase in $\kappa$ per added noise is 0.01 per G
for MDI and 0.02 per G for SP and HMI. The length of the
plateau and steepness of the increase are related to the inherent
noise in the data: The less instrumental noise there is, the
more sensitive the cancellation function is to added noise. SP,
which has the lowest nominal noise level, has the shortest plateau and
steepest increase, while MDI, the noisiest instrument, is less
sensitive to noise. Consistent with the noise experiment, SP has the
smallest measured exponents and MDI the largest, suggesting that MDI
magnetograms are more strongly affected by noise.

\begin{figure*}
\includegraphics[width=10cm]{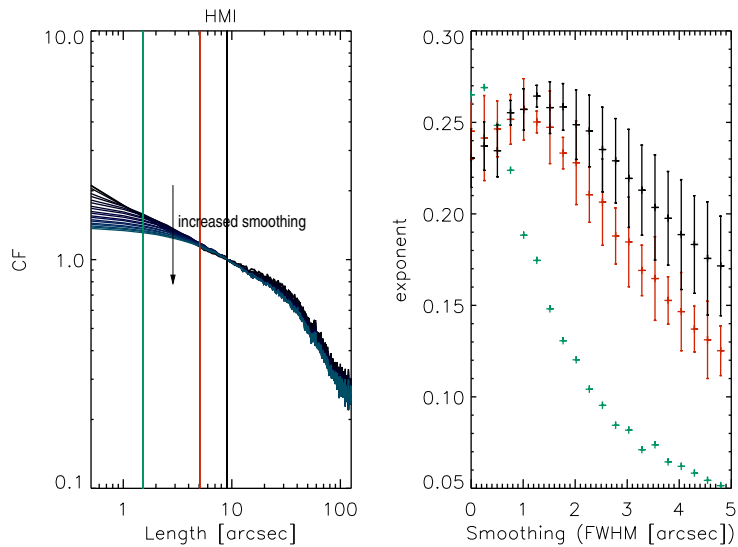}
\caption{Effects of spatial smoothing. Left: HMI cancellation
functions computed with increased spatial smoothing (Gaussian
smoothing with full width at half maximum (FWHM) increasing in
increments of 0.5 pixels). Right: Exponents as a function of Gaussian
FWHM (\ks\ in green, \ki\ in red and \kl\ in black).}
\label{fig:hmi-smooth}
\end{figure*}

Figure \ref{fig:hmi-smooth} shows how spatial smoothing changes the cancellation function. The effect of
smoothing is opposite to that of increasing the noise (recall that
$\kappa=0$ for a smooth field and $k=1$ for noise). The exponents
decrease with increased spatial smoothing.   {This demonstrates}
how seeing conditions and spatial over-sampling can  {reduce} the
measured exponents.

\section{Conclusions}

{Significant artifacts in measuring the cancellation function from
 current observations have been identified. The cancellation exponent
 is sensitive to noise level differences between instruments: Noise
 increases both $\kappa$ and the scatter between individual
 measurements of $\kappa$. Not surprisingly, $\kappa$ decreases by
 seeing and spatial over-sampling.  We also find that $\kappa$ is
 sensitive within a single instrument's observations to magnetic flux
 imbalances and mean unsigned flux.  A simple model suggests that this
 dependence is the effect of network elements: They reduce $\kappa$.}
This simple model also explains why the value of 
$\kappa$ calculated in PGDS2009 and \inlinecite{St2011} differ: The
network is enhanced in the latter study, decreasing $\kappa$.

{Of all the observational artifacts, the effect of the network can
 be removed by masking network elements} out of magnetograms.  We
then find values of $\kappa_I\approx0.58$ for HMI and
$\kappa_I\approx0.38$ for SP. (For MDI the removal of network was not
fully successful and the dependence of \e\ on the presence of network
elements was not entirely removed.)  The difference in $\kappa$ between HMI and SP is due to HMI magnetograms being demonstrably more noisy:
HMI's noise level is $\approx 4\,$G and SP's is $\approx 2\,$G, at
least for determinations of the cancellation exponent (see Figure
\ref{fig:noise}).  Since noise first affects the smallest scales of
the cancellation function, high resolution (diffraction limit,
pixel size, and seeing) observations with minimal noise are needed to measure the
exponent on scales below a couple of arcseconds.   {Therefore, no
 conclusive value of the unsigned flux can be given by cancellation
 function analysis with current instruments.}

{An upper bound of the unsigned flux is the best that can be done
 with the available data.  Since SP has the lowest noise level, we
 can safely conclude $\kappa\lessapprox0.38$.  The quiet-Sun magnetic
 field is not a completely self-similar field: It will become smooth
 over an order-of-magnitude in scales near the magnetic diffusion
 scale} (below which there is no further cancellation).
{However, using $\kappa\lessapprox0.38$ and that the power-law
 ends at $l_0 \ge 80\,$m} (the magnetic diffusion scale estimate from
PGDS2009)  {in Equation (\ref{eq:extrapolate}), we have an upper
 bound on the true mean unsigned flux.  We find that}
$\langle|B_z|\rangle_{\mathrm{true}}\equiv\mathrm{Flux}_0\lessapprox268\,$G
(standard deviation 44 G) from SP data with exposure times longer than
5 s.   {(Note that if the power-law for some reason ends at even
 larger scales than the $\approx800\,$m beginning of the diffusive
 range, this bound still holds.)}  The noisier instruments, MDI and
HMI, give significantly higher values. Both the noise level and
sensitivity of the instruments affect the estimates. SP being the
least noisy of the instruments can be considered as an upper bound for
the flux.    {\inlinecite{MaSaLaDeTrBu2006} found a similar upper bound,
 $\langle|B_z|\rangle_{\mathrm{true}}\lessapprox200-300\,$G from the
 scattering line polarization of Ce {\sc ii}.}

We found no correlation between $\kappa$ and global mean unsigned flux: The
small scale ($<$10\arc) field distribution/fractal geometry is not
affected by the global magnetic configuration, at least not during a
rising phase of the cycle sampled by HMI. {A study of HMI cancellation functions from the end of the previous solar minimum past the next solar maximum may show how flux from decaying active regions affects the flux distribution in the quiet Sun, and possibly give indications of the relative importance of possible quiet-Sun small-scale dynamo action in different phases of the solar activity cycle.}  

{It should be emphasized that the present analysis and findings are for quiet Sun magnetic fields which are known to be turbulent. Issues identified in the analysis most likely do not affect as strongly measurements of active region fractal dimensions.}   
The significance of noise in the cancellation statistics may extend to
the studies of fractal dimensions and flaring probability
({\it e.g.}, \opencite{Georgoulis2011}). Based on the current analysis, using
less-noisy magnetograms, such as HMI, may show the fractal dimension {of the magnetic flux} to
still have some predictive capability for flaring active regions. To
establish the usability of fractal analysis for flare forecasting,
however, a large sample of flaring and non-flaring
active region magnetograms with little noise is needed. {A statistical study of flaring and non-flaring HMI vector magnetograms could also be used to address the differences of using line of sight flux and current helicity for predictions.}

\begin{ack}
 JPG gratefully
acknowledges the support of the U.S. Department of Energy through the
LANL/LDRD Program for this work. {\it Hinode} is a Japanese mission developed and launched by ISAS/JAXA, with NAOJ as domestic partner and NASA and STFC (UK) as international partners. It is operated by these agencies in co-operation with ESA and NSC (Norway) Data provided by the SOHO/MDI consortium. SOHO is a mission of international cooperation between ESA and NASA. SDO is a mission for NASA's Living With a Star program.
\end{ack}
\section*{Appendix}

\begin{figure*}
\includegraphics[width=12.cm]{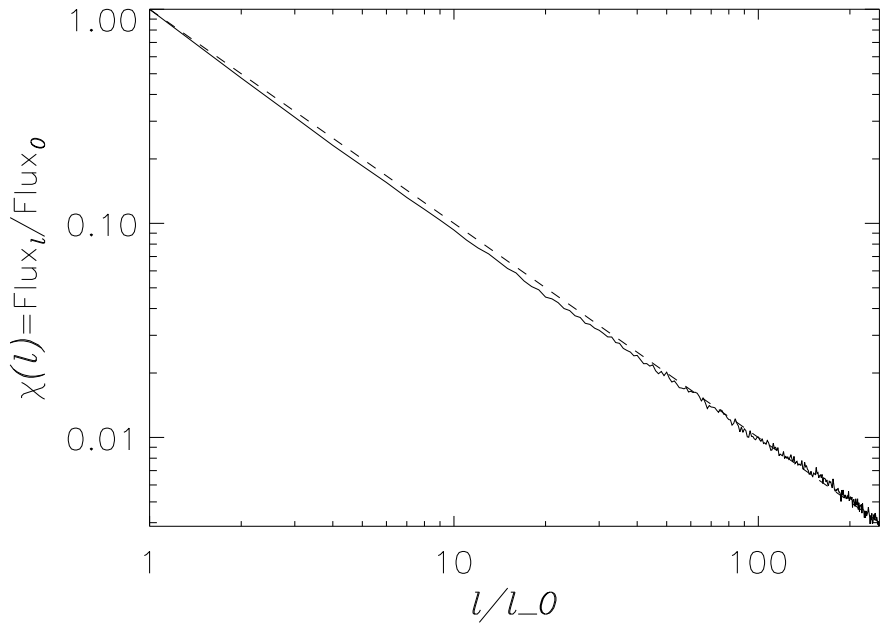}
\caption{ {Ratio of net flux to unsigned flux $\chi(l)$ versus scale ($l$) for noise
with correlation length $l_0$.
The ratio is given by  $\frac{l_0}{l}$ (dashed line).}}
\label{fig:kappais1}
\end{figure*}

{For a
completely self-similar field, knowledge of net flux at any scale and
of the power-law scaling exponent implies knowledge of net flux at all
scales and, hence, the unsigned flux.  This is most
 easily seen in the example of a vertical random field [$f_z$]. 
 For a
random field, the integral over an area is proportional to the square
root of the area (random walk),
\begin{equation}
\bigg{|}\int_{\mathcal A_i(l)\in\mathcal A}f_z\mathrm{d}a\bigg{|} \propto l\,,
\end{equation}
while the integral over a strictly positive field is proportional to its area,
\begin{equation}
\int_{\mathcal A_i(l)\in\mathcal A}|f_z|\mathrm{d}a \propto l^2\,.
\end{equation}
Therefore,
\begin{equation}
\chi(l)\propto \frac{1}{l}\,,
\end{equation}
and we identify $\kappa=1$ for pure noise.
For scales smaller than the correlation length of the noise, the net
flux is equal to the unsigned flux,
\begin{equation}
\mathrm{Flux}_{l} = \mathrm{Flux}_{0} \qquad \forall l \le l_0\,,
\label{eq:kappais1a}
\end{equation}
and for larger scales,
\begin{equation}
\mathrm{Flux}_{l} = \frac{C}{l} \qquad \forall l \ge
l_0\,.
\label{eq:kappais1b}
\end{equation}
The constant $C$ is determined by equating Equations (\ref{eq:kappais1a})
and (\ref{eq:kappais1b}) at the correlation length, $l=l_0$:
\begin{equation}
\frac{\mathrm{Flux}_{l}}{\mathrm{Flux}_{0}} = \frac{l_0}{l} \qquad \forall l \ge
l_0\,.
\label{eq:kappais1c}
\end{equation}
This power-law dependence, Equation. (\ref{eq:kappais1c}), is shown in Figure
\ref{fig:kappais1}.  This means that the unsigned flux can be exactly
calculated given the net flux at any scale and the scale at the end of
the power-law, $l_0$,
\begin{equation}
\mathrm{Flux}_{0} = \mathrm{Flux}_{l}\frac{l}{l_0}\,,
\end{equation}
for a random field.
For any other purely self-similar field with
cancellation exponent $\kappa$,
\begin{equation}
\mathrm{Flux}_{l} = \mathrm{Flux}_{L}\big{(}\frac{L}{l}\big{)}^\kappa \qquad \forall l,L \ge l_0\,.
\label{eq:extrapolate}
\end{equation}
For the extreme case of a unipolar field, $\kappa=0$ and
$\mathrm{Flux}_{l} = \mathrm{Flux}_{0}$ for all scales $l$.  In general, $0\le\kappa\le1$. }

%

\end{article} 
\end{document}